\documentclass[twocolumn, tighten, times]{aastex61}
\usepackage{graphicx}
\usepackage{epstopdf}
\usepackage{subfigure}
\usepackage{color}
\usepackage{soul}
\usepackage{tabularx}
\usepackage{booktabs}

\newcommand{\ra}[1]{\renewcommand{\arraystretch}{#1}}

\renewcommand\baselinestretch{1.3}

\shorttitle{The environment of massive green valley galaxies at $0.5<z<2.5$}
\shortauthors{Chang et al.}

\begin{document}

\title{The physical properties of massive green valley galaxies as a function of environments \\ at $0.5<z<2.5$ in 3D-\textit{HST}/CANDELS fields}


\author{Wenjun Chang}
\affil{Institute of Astronomy and Astrophysics, Anqing Normal University, Anqing 246133, China; wen@mail.ustc.edu.cn}
\affil{Department of Physics and Astronomy, University of California, Riverside, 900 University Avenue, Riverside, CA 92521, USA}
\affil{Department of Astronomy, University of Science and Technology of China, Hefei 230026, China; xkong@ustc.edu.cn}
\affil{School of Astronomy and Space Sciences, University of Science and Technology of China, Hefei 230026, China}

\author{Guanwen Fang}
\affil{Institute of Astronomy and Astrophysics, Anqing Normal University, Anqing 246133, China; wen@mail.ustc.edu.cn}

\author{Yizhou Gu}
\affil{Department of Astronomy, School of Physics and Astronomy, Shanghai Jiao Tong University, Shanghai 200240, China}

\author{Zesen Lin}
\affil{Department of Astronomy, University of Science and Technology of China, Hefei 230026, China; xkong@ustc.edu.cn}
\affil{School of Astronomy and Space Sciences, University of Science and Technology of China, Hefei 230026, China}

\author{Shiying Lu}
\affil{School of Astronomy and Space Science, Nanjing University, Nanjing 210093, People's Republic of China}
\affil{Key Laboratory of Modern Astronomy and Astrophysics (Nanjing University), Ministry of Education, Nanjing 210093, China}

\author{Xu Kong}
\affil{Department of Astronomy, University of Science and Technology of China, Hefei 230026, China; xkong@ustc.edu.cn}
\affil{School of Astronomy and Space Sciences, University of Science and Technology of China, Hefei 230026, China}

\begin{abstract}
To investigate the effects of environment in the quenching phase, we study the empirical relations for green valley (GV) galaxies between overdensity and other physical properties (i.e., effective radius $r_{\rm e}$, S\'{e}rsic indices $n$, and specific star formation rate sSFR). Based on five 3D-{\it HST}/CANDELS fields, we construct a large sample of 2126 massive ($M_{\star} > 10^{10} M_{\sun}$) GV galaxies at $0.5<z<2.5$ and split it into the higher overdensity quarter and the lower overdensity quarter. The results shows that GV galaxies in denser environment have higher $n$ values and lower sSFR at $0.5< z <1$, while there is no discernible distinction at $1 < z < 2.5$. No significant enlarging or shrinking is found for GV galaxies in different environments within the same redshift bin. It suggests that a dense environment would promote the growth of bulge and suppress star formation activity of GV galaxies at $0.5< z <1.5$, but would not affect the galaxy size. We also study the dependence of the fraction of three populations (Blue Cloud, Green Valley, and Red Sequence) on both environments and $M_{\star}$. At a given $M_{\star}$, blue cloud fraction goes down with increasing environment density, while red sequence fraction is opposite. For the most massive GV galaxies, a sharp drop appears in the denser environment. Coupled with the mass dependence of three fractions in different redshift bins, our result implies that stellar mass and environments jointly promote the quenching process. Such dual effect is also confirmed by re-calculating the new effective GV fraction as the number of GV galaxies over the number of non-quiescent galaxies.
\end{abstract}


\section{Introduction}\label{sec1:intro}

A flood of observations on large samples of galaxies found evident cessation of star formation activities during the evolution of galaxies. The cessation of star formation is widely known as ``quenching'' and is assumed to produce passive galaxies in which the star formation rate (SFR) is very low, commonly resulting in the bimodal distribution of galaxies. In the color--magnitude diagram, the narrow red peak, in which abundant quiescent galaxies (QGs) and a small amount of dusty star-forming galaxies (SFGs) are distributed along with a linear sequence (\citealt{Blanton+2009}), is normally called ``red sequence'' (RS). The blue peak occupying an extended region mainly consists of SFGs (\citealt{Kauffmann+03}), similarly called ``blue cloud'' (BC). An intermediate zone between BC and RS is commonly known as ``green valley'' (GV), initially proposed and described in several GALEX papers (e.g., \citealt{Martin+2007, Salim+07, Wyder+2007}).

GV galaxies (see the review of \citealt{Salim+14}) are often thought to be in a transitional phase. The existence of GV galaxies argues for a continuum of properties from SFGs to QGs (\citealt{Wyder+2007,Salim+14}). Over the past decades, mechanisms of galaxy quenching remains an unsolved problem. Many previous researches emphasized the role of stellar mass ($M_{\star}$) and environment in the cessation of star formation, generally referred to as ``mass quenching'' (\citealt{PengYJ+10, Somerville+2015, Penny+2018}) and ``environment quenching'' (\citealt{PengYJ+10, Darvish+2015}). A plausible explanation of ``mass quenching'' is that the feedback from active galactic nucleus (AGN) plays an important role in regulating star formation and in quenching galaxies (\citealt{Hopkins+2005, Hopkins+2014, Somerville+2015, Penny+2018}). At the same time, a number of authors have considered that the external environment might be the key factor of suppressing the star formation via some mechanisms to consume the gas reservoir, such as major/minor merger (\citealt{Springel+05}), ram pressure stripping (\citealt{GG+72}), and harassment (\citealt{F&S+81}). High environmental density might subsequently increase dust attenuation (e.g., \citealt{Koyama+2013, Sobral+2016}) and enrich interstellar medium of SFGs (\citealt{Sobral+2015, Darvish+2015}).

It is noteworthy that galaxy morphology varies with the environment. Generally, galaxies in the lower density environments (field) are bluer, more star-forming, and more disc-like, while galaxies in higher density environments (cluster) are older, redder, less star-forming, and more elliptical (\citealt{Dressler+1984, Kauffmann+04}). \citet{Dressler+1980} found that there is a definite relationship between local galaxy density and morphological type at $z < 0.06$, which is further confirmed in other studies (\citealt{Guzzo+1997, Goto+2003, Fasano+2015}). The dependence of galaxy morphology on the environment appears not only in the local Universe but also at intermediate and high redshifts ($z\sim1-2$, e.g., \citealt{Dressler+1997, Vanderwel+2007, Allen+2015, Allen+2016}). \citet{Allen+2015} compared the morphology of SFGs and QGs in different environments (cluster vs. field). They found that cluster SFGs had higher S\'{e}rsic indices ($n$) than field SFGs at $0 < z < 2$, while there is no difference in light profile and galaxy size for QGs between field and cluster environments. However, it is not consistent with the results of others that QGs in the cluster have shallower profiles (lower $n$) and larger sizes (\citealt{Bassett+2013, Strazzullo+2013, Yoon+2017}).

Considering that the environment affects both star formation activity and morphology in galaxies at the transitional stage, GV galaxies could be one of the suitable samples to study the environmental effect on physical properties. In our previous series of work, \citet{Gu+18} have assembled a large sample of massive ($M_{\star} > 10^{10}~M_\odot$) RS, GV, and BC galaxies at $0.5\leqslant z \leqslant 2.5$ in five fields of 3D-{\it HST}/CANDELS, and investigated their morphology, dust content, and environments and revealed a mass-dependent ``downsizing'' quenching picture. Then, \citet{Gu+2019} analyzed the mass dependence of morphology and star formation activity for three populations. We found that the structural properties of GV galaxies are intermediate between those of BC and RS galaxies at fixed stellar mass bins at $z<2$, and both GV and BC galaxies have similar sizes and compactness at the high-mass end. It implies that GV galaxies could go through a morphological transformation of bulge buildup at $z<1.5$, which is consistent with our results of \citet{Lu+2021}. We found the effect of the morphological quenching mechanism on star formation activity at $0.5< z <2.5$, after eliminating any possible AGN candidates and considering the stellar mass influence. But we had not studied the effect of the environment on the physical properties of GV galaxies until \citet{Gu+21} defined a dimensionless overdensity (1+$\delta^{'}$) as the environmental indicator by adopting a Bayesian method to consider the contributions of all the $N$th nearest neighbors. Based on this improved environmental method, in this work we set out to explore empirical relations between galaxy environments and other physical properties, including parametric and non-parametric structure, and star-forming parameters, for GV galaxies at $0.5< z <2.5$. To understand the quenching process for massive galaxies, we also analyze the transformation of different fractions for RS, GV, and BC galaxies.

The structure of our paper is organized as follows.
Data and sample selection are described in Section~\ref{sec2:DS}.
We present the environmental dependence of structural parameters and sSFR in Section~\ref{sec3:EDP}. Section~\ref{sec4:FQG} contains our analysis about the different fractions for RS, GV, and BC populations and the distribution of effective fractions for GV galaxies. Finally, a summary is given in Section~\ref{sec5:Sum}. Throughout our paper, we adopt the cosmological parameters as following: $H_0=70\,{\rm km~s}^{-1}\,{\rm Mpc}^{-1}$, $\rm \Omega_m=0.30$, and $\Omega_{\Lambda}=0.70$. All magnitudes adopted in this paper are in the AB system.

\section{Data and Sample selection} \label{sec2:DS}

\subsection{3D-{\it HST} and CANDELS} \label{sec2.1:C3D}
The 3D-{\it HST} and CANDELS  programs cover over $\sim$900 arcmin$^{2}$ in five different fields: AEGIS, COSMOS, GOODS-N, GOODS-S, and UDS, observed by a number of space-based and some ground-based telescopes. It thus provides abundant imaging data from ultraviolet (UV) to infrared (IR) bands via the high-quality WFC3 and ACS spectroscopy and photometry \citep{Grogin+11, Koekemoer+11, Skelton+14, Momchheva+16}. In the 3D-{\it HST}/CANDELS, many physical properties of galaxies have been measured by utilizing the homogeneous multi-band data, including the parameters of the stellar population (\citealt{Skelton+14, Whitaker+14, Momchheva+16}) and structure (\citealt{vdW+14}).

In this work, redshifts and rest-frame colors are taken from \citet{Momchheva+16}. It is an updated version of the photometric catalog from \cite{Skelton+14} with the grism redshifts from the fits of G141 grism spectroscopy. We refer to \citet{Momchheva+16} for the full details. Following \citet{WangT+17}, stellar mass and dust attenuation are re-estimated based on the stellar population synthesis models from \citet{Ma+05} via the FAST code \citep{Kriek+09}. The \citet{Ma+05} models are known to give a better description of stellar populations for high-redshift SFGs whereby taking the contribution of the thermally pulsating asymptotic giant branch stars into account. We assume an exponentially declining star formation history with an e-folding time $\tau \sim 0.1-10$ Gyr, a \cite{Kroupa+01} initial mass function (IMF) and solar metallicity. The dust attenuation ($A_V$) varies from 0 to 4 in steps of 0.1 following the \cite{Calzetti+00} law. It is worthy of mentioning that we prefer to take the spectroscopic (or grism) redshifts if available. Otherwise, we will use the photometric redshifts from \citet{Skelton+14} instead.

\begin{figure*}
\centering
\includegraphics[scale=0.8]{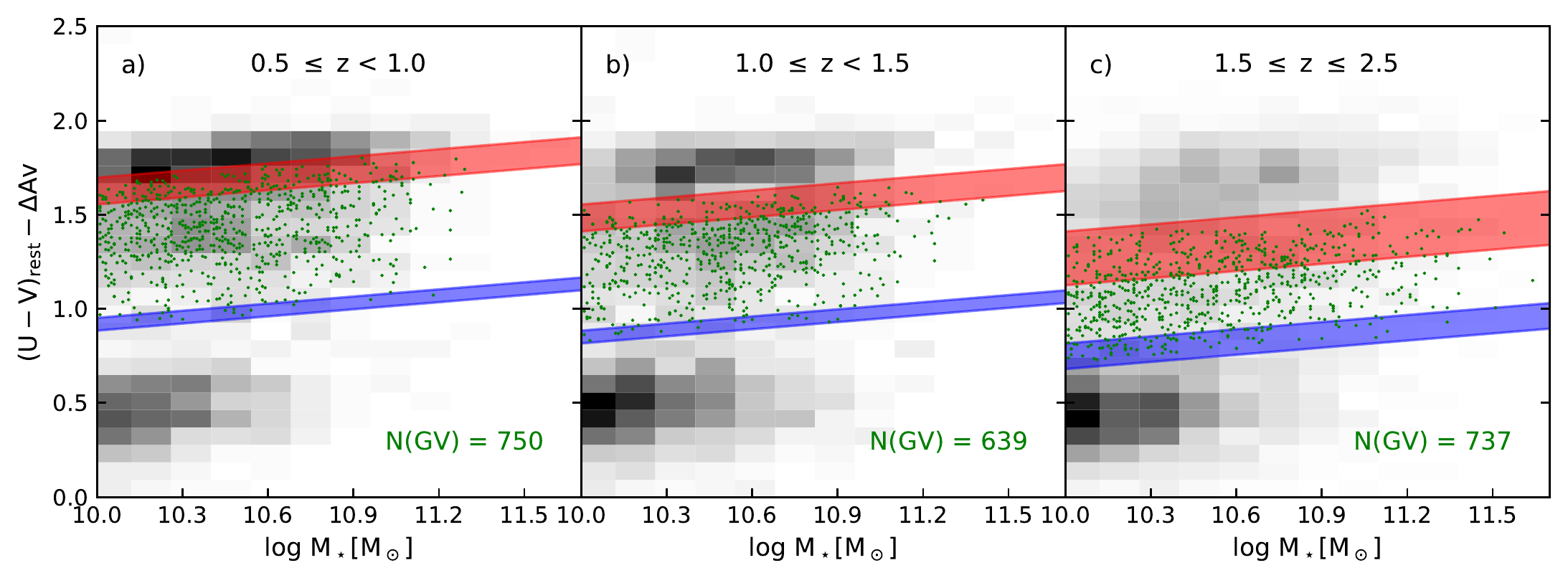}
\caption{The extinction-corrected rest-frame U-V color of GV galaxies as a function of stellar mass in three redshift bins. In each panel, the background grayscale shallows represent the distribution for all our parent massive galaxies. Green points are GV galaxies. The two criteria for each redshift bin from \citet{WangT+17} are separately shown as red and blue bands, where the edge of the ribbon corresponds to the range of redshift. The number of GV galaxies in each redshift bin has been shown in the bottom of each panel.
\label{fig:f01_sample}
}
\end{figure*}

\subsection{Structural Parameters} \label{sec2.2: Structure}

Quantitative morphological and structural analysis has been developed and complemented in the past decades. There are two main measurements of morphology: parametric modelling of surface brightness profile (\citealt{Sersic1968, PengCY+02, PengCY+10}) and non-parametric morphology (\citealt{Abraham+94, Lotz+04, Conselice+14}).

Parametric models are important in describing light profile to estimate the galaxy size and bulge-to-disk decomposition (\citealt{Buitrago+08, Wuyts+11a, Lang+14, vdW+14}).  S\'{e}rsic indices ($n$) and effective radius ($r_{\rm e}$) in CANDELS fields are taken from \cite{vdW+14} which have been measured using GALFIT (\citealt{PengCY+02}). 3D-{\it HST} catalog has been matched with J band and H band of CANDELS by Rainbow Database\footnote{\url{https://arcoirix.cab.inta-csic.es/Rainbow_Database/Home.html}}. The rest-frame optical morphologies are traced by J-band (F125W) imaging at $0.5< z<1.5$ and by H-band (F160W) imaging at $1.5<z<2.5$ in this work.

Non-parametric methods are usually used to identify irregular structures in galaxies, especially for high-redshift galaxies, with high irregularity (\citealt{Bluck+12, Conselice+14}). We have performed the measurements of the Gini coefficient ($G$; \citealt{Lotz+04}) and the second-order moment of the 20\% brightest pixels ($M_{20}$; \citealt{Lotz+04}) on the NIR images using the \textsc{Morpheus} software developed by \cite{Abraham+07}. More detailed definitions and calculations for these non-parametric measurements refer to \citet{Gu+18}. The effect of signal-to-noise ratio (S/N) on the measurement of non-parametric parameters has been tested by previous work \citep{Kong+09, Wang+12, Fang+15, Gu+18, Gu+2019, Lu+2021}. In sample selection (see Section~\ref{sec2.5: sample select}) we adopt a flag of $\tt use\_phot = 1$ that ensures a reliable detection in $H$-band (F160W) with S/N$>$3 and thus reliable non-parametric measurements.

\subsection{Star Formation Rates} \label{sec2.3:SFR}

Young and massive stars produce a large amount of UV photons, which are partly absorbed by interstellar dust and re-emit in IR. Thus, accurate SFR estimation should include both unobscured and obscured parts. The unobscured SFR can be derived from the observed rest-frame UV luminosity, while the obscured one can be estimated through a single mid-infrared measurement, e.g., MIPS 24 $\mu$m \citep{Battisti+2015,Lin+2016}.

In this work, we employ the total SFR by combining UV and IR emissions, provided by \citet{Whitaker+14}. Considering the \cite{Bell+05} conversion and the \cite{Kroupa+01} IMF, the $\rm SFR_{UV+IR}$ can be estimated by:
\begin{equation}
{\rm SFR_{UV+IR}}[M_\sun~{\rm yr}^{-1}]=9.8\times10^{-11}(L_{\rm IR}+2.2L_{\rm UV})/L_{\sun},
\end{equation}
where $L_{\rm UV}=1.5 \times L_{\rm 2800}$ is estimated by the rest-frame continuum luminosity at 2800\AA, and $L_{\rm IR}$ is the integrated luminosity at 8$\sim$1000 $\mu$m converted from the Spitzer/MIPS 24 $\mu$m data using a single luminosity-independent template (\citealt{Franx+08, Wuyts+2008}).

If the MIPS 24 $\mu$m data are unavailable, we assume the dust attenuation curve from \cite{Calzetti+00} and correct the effect of dust attenuation on $\rm SFR_{UV}$. In this case, the corrected $\rm SFR_{UV}$ can be derived as follows:
\begin{equation}
 {\rm SFR_{UV, corr}}[M_\sun~{\rm yr}^{-1}]={\rm SFR_{UV}}\times10^{0.4\times 1.8 \times A_V},
\end{equation}
where ${\rm SFR_{UV}}=3.24 \times 10^{-10} \times  L_{2800}/L_{\sun}$. 
The factor of 1.8 converts $A_V$ to that at 2800\AA\ when adopting the \cite{Calzetti+00} attenuation curve.

\subsection{Measurements of Environmental Overdensities} \label{sec2.4:MED}
The different definitions of environmental density describe disparate physical properties on different physical scales (\citealt{Muldrew2012, Etherington+15}).
In the paper, we adopt the dimensionless overdensity as the indicator of the relative local environment of galaxies,  described in detail in \citet{Gu+21}.
A magnitude-limited sample is selected for the measurements of environmental overdensities. This sample mixes galaxies with spectroscopic, grism, or photometric redshifts and is selected by the following criteria: (1) H-band apparent magnitudes
$ \rm F160W < 25$ that guarantees the uncertainty of photometric redshifts $\sigma_z = 0.02$, (2) a redshift range of $0.5 < z < 2.5$, and (3) flag {$\tt use\_phot = 1$} that ensures reliable photometries of galaxies (see Section \ref{sec2.5: sample select} for more details). About 30\% of galaxies in this sample with spectroscopic or grism redshifts provide a more convinced derivation of galaxies environments, although using purely photometric redshifts to construct the overdensities would not change our results in general. In the following,  we summarise the relevant information on the environmental indicator.

The Bayesian metric in \citet{Cowan+2008} is adopted to estimate the local environmental density, which is defined as $\Sigma^{'}_{N}$ $\propto$ 1/($\Sigma^{N}_{i=1} d_i^2$). Here $d_i$ is the projected distance of the $i$th nearest neighbor in the projected two-dimensional space within the individual redshift slice. This Bayesian environmental density considers the contribution of all $N$th nearest neighbors, which can improve accuracy in mapping the probability density distribution compared to the traditional method (\citealt{2005AJ....129.1096I}). Given that photometric redshifts have large uncertainties, we choose the redshift slice as $ \left|\Delta z\right| =  2 \sigma_{z} (1+z)$, where $\sigma_{z} = 0.02$ is the typical uncertainty of photometric redshifts of our sample, and $z$ is the redshift of target galaxy to estimate the environment.

With increasing redshift, the comoving number densities of the observed galaxies are expected to decrease. Thus, we adopt a dimensionless overdensity of \citet{Gu+21}, 1+$\delta^{'}_{N}$, as the indicator of galaxy environment:
\begin{equation}
\label{eq: density}
1+\delta^{'}_{N} = \frac{\Sigma^{'}_{N}}{\langle \Sigma^{'}_{N} \rangle_{\rm uniform}} = \frac{\Sigma^{'}_{N}}{k^{'}_{N}  \Sigma_{\rm surface}},
\end{equation}
where $\langle \Sigma^{'}_{N} \rangle_{\rm uniform}$ is the standard value of Bayesian density when galaxies are distributed in the uniform environment. At a given density $\Sigma_{\rm surface}$, $\langle \Sigma^{'}_{N} \rangle_{\rm uniform}$ can be calculated by a linear correction coefficient $k^{'}_{N}$
(see also in the Appendix of \citealt{Gu+21}). Thus, $\log(1+\delta^{'}_{N}) >$ 0 indicates that the environmental density of a galaxy exceeds the density standard in the uniform condition and vice versa.

Due to the Poisson noise and the possible contamination of foreground and background galaxies,
a small value of $N$ may cause fluctuation in the density values.
We therefore adopt the overdensity based on the distances to all 10 nearest neighbors ($\Sigma^{'}_{10}$) as the indicator of galaxy environments in this work. statistically

\subsection{Sample Selection} \label{sec2.5: sample select}
Based on the magnitude-limit sample, we focus on the massive galaxies with $\log (M_{\star}/M_\sun) \geqslant 10$.
The flag of $\tt use\_phot = 1$ is set to  choose galaxies with a reliable detection
following these criteria: 1) not a star, or bright enough to be a reliable galaxy; 2) not close to a bright star; 3) well-exposed in the F125W and F160W bands; 4) a S/N of f$\_$F160W / e$\_$F160W $>$ 3; 5) has a passable photometric redshift fit; 6) has a ``non-catastrophic'' stellar population fit, with $\log M_{\star}>$ 0 \citep{Skelton+14}. To ensure reliable measurement of the structural parameters, we also make an additional cut on {\it H}-band, H$_{\rm F160W} <$ 24.5, which could get rid of faint galaxies and guarantee much higher reliable photometric redshifts (\citealt{Momchheva+16}). Finally, the parent sample contains 7850 massive galaxies at 0.5 $<$ z $<$ 2.5 from five 3D-{\it HST}/CANDELS fields, about 58\% of galaxies have spectroscopic or grism redshifts. Considering about half of sample with photometric redshifts, the bias from galaxies spectroscopic redshifts which trend to be brighter might be relieved.

Given that the intrinsic rest-frame colors depend on stellar mass and redshift, we use the extinction-corrected rest-frame U-V colors to divide galaxies into RS, GV, and BC populations as well, as \citet{WangT+17} did. The separation criteria are as follows:
\begin{eqnarray} \nonumber
 (U - V )_{\rm rest} - \Delta A_V = 0.126 \log (M_{\star}/M_{\sun}) + 0.58 - 0.286z; \,\, \nonumber  \\
\nonumber
 (U - V )_{\rm rest} - \Delta A_V = 0.126 \log (M_{\star}/M_{\sun})  - 0.24 - 0.136z, \,\, \nonumber
\end{eqnarray}
where $\Delta A_V=0.47 \times A_V$ is the extinction correction of rest-frame U-V color, the value of 0.47 is the correction factor from the \cite{Calzetti+00} attenuation law. Then we obtain 2566 RS galaxies, 2126 GV galaxies, and 3158 BC galaxies.

In Figure~\ref{fig:f01_sample}, we show the extinction-corrected rest-frame U-V color of GV galaxies as a function of stellar mass in three redshift bins from 0.5 to 2.5. The criteria for each redshift bin from \citet{WangT+17} are separately shown as red and blue bands, where the width of the ribbon corresponds to the range of redshift. GVs are distributed below the upper limit of the first separation and above the lower limit of the second separation. The background grayscales represent the distributions for all massive galaxies in our parent sample, and also show the separation for BC and RS. The number of GV in each bin is also shown in each panel.
Due to the redshift-dependence of the GV definition, we can find that high-redshift galaxies tend to be bluer than low-redshift galaxies in Figure\ \ref{fig:f01_sample}.

\section{Environmental Effect on Physical Properties} \label{sec3:EDP}

In this section, we discuss the effect of the environment on different physical properties of GV galaxies, including parametric ($n$ and r$_e$), non-parametric ($G$ and M$_{20}$) structures, and sSFR. Considering that the dominated quenching mechanism might vary across cosmic time (\citealt{Iovino+2010}), we carry out the research in three redshift bins (0.5 $\leq z <$ 1.0, 1.0 $\leq z <$ 1.5, and 1.5$\leq z \leq$ 2.5). In each redshift bin, GV galaxies are divided into four equal bins according to the local overdensity 1+$\delta^{'}_{\rm 10}$. In the following, we compare the physical properties of GV galaxies in the highest and lowest overdensity bins, namely the highest and lowest density quarters.
To qualify the correlation between stellar mass and different physical properties and the corresponding differences between the highest and lowest density quarters in three redshift bins, we calculate the Spearman correlation coefficients and perform 2-Dimensional Kolmogorov–Smirnov (2D-KS) test (\citealt{Peacock1983, Fasano1987}), respectively. It is assumed that there is a significant difference between two density quarters when the p-value of 2D-KS test is smaller than 0.05. All results are listed in Table\ \ref{tab: corr_KS}.

\begin{figure*}
\includegraphics[width=\textwidth]{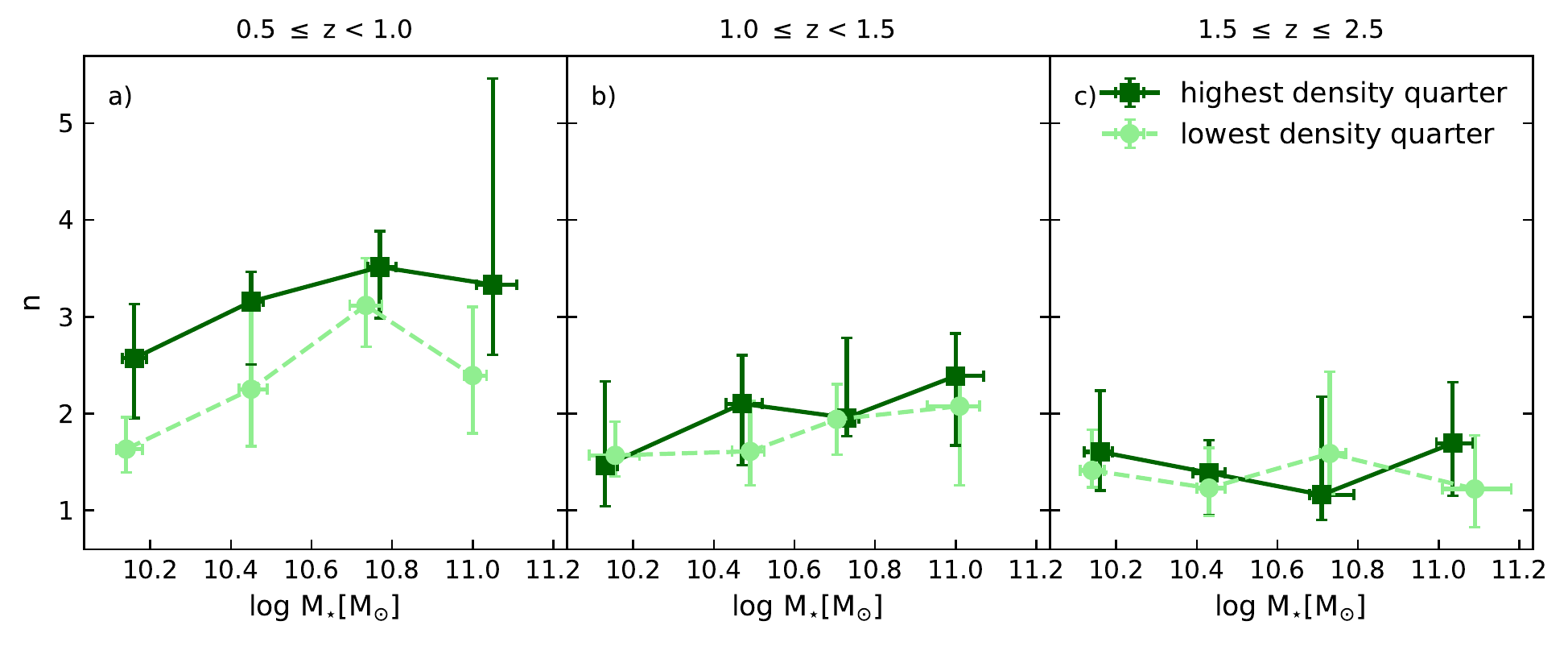}
\includegraphics[width=\textwidth]{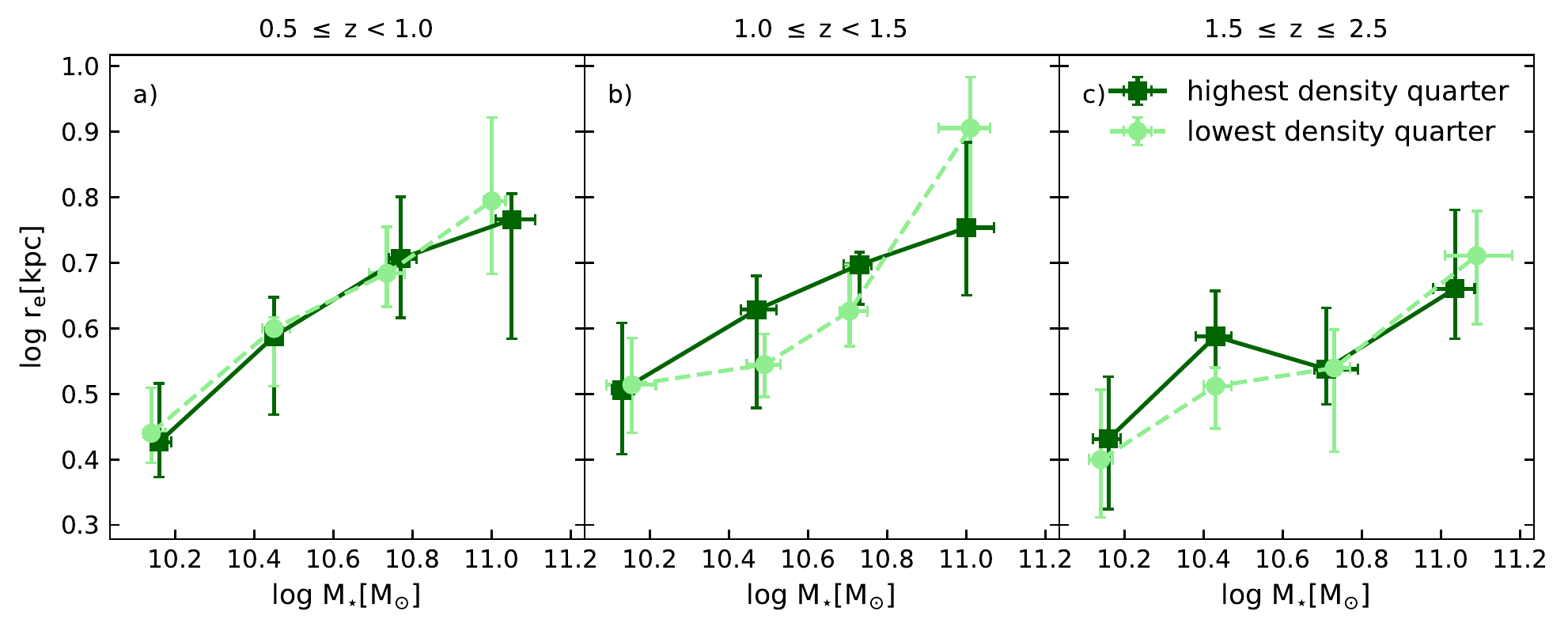}
\caption{S\'{e}rsic index $n$ (top panels) and effective radius r$_e$ (bottom panels) as functions of $M_{\star}$ in the highest (dark green) and lowest (light green) density quarters. The error bars are the 2$\sigma$ uncertainties drawn from the bootstrap method with the 1000 times resamples.}
\label{fig02-all}
\end{figure*}

\begin{figure*}
\includegraphics[width=\textwidth]{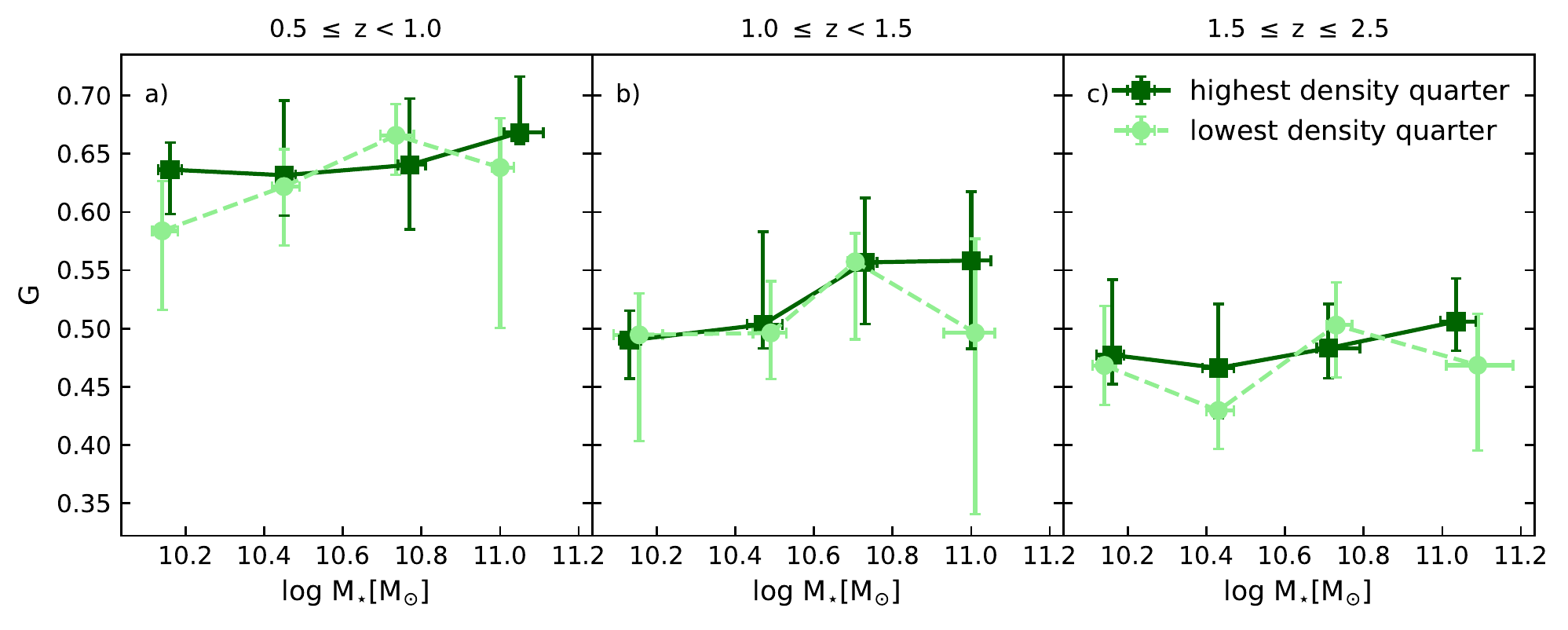}
\includegraphics[width=\textwidth]{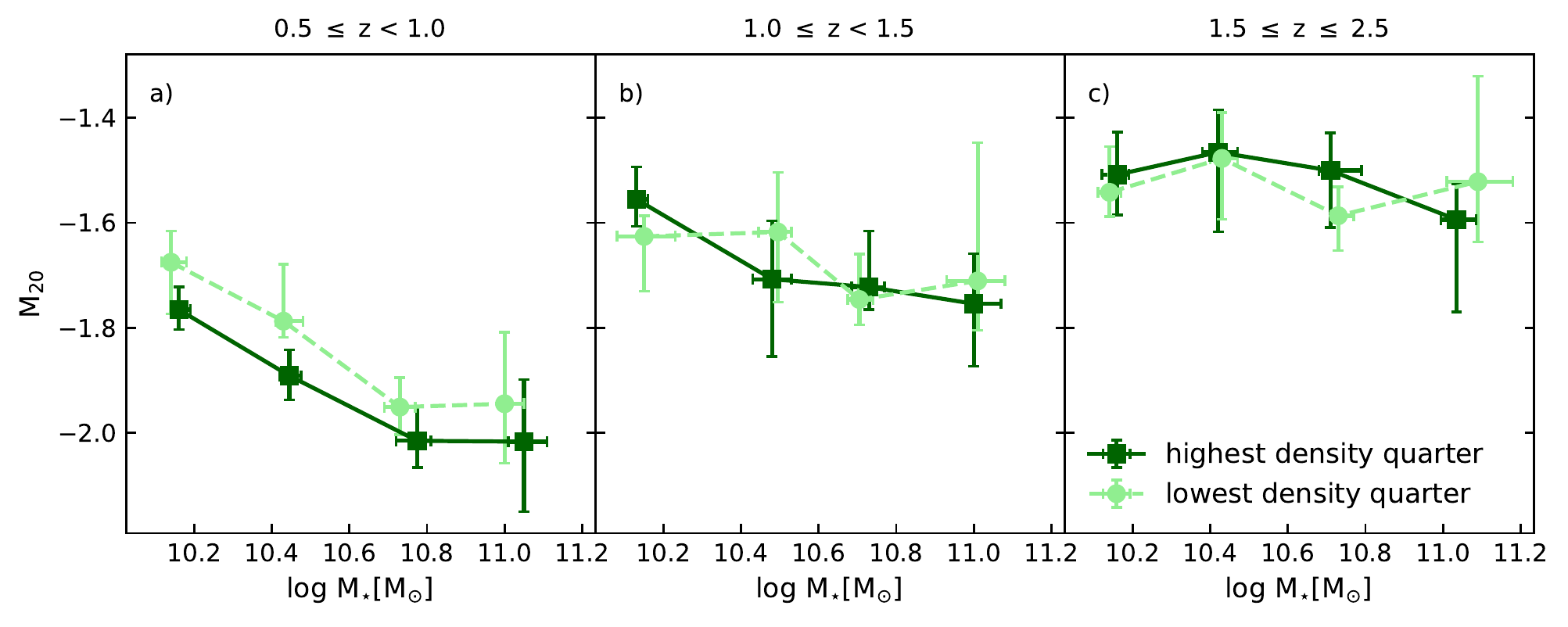}
\caption{Gini coefficient $G$ (top panels) and the second order 20\% bright pixels $M_{20}$ (bottom panels) as functions of $M_{\star}$ in two extreme environments. The colors and markers are same as Figure~\ref{fig02-all}. }
\label{fig03-G-M20}
\end{figure*}

\subsection{Parametric Structures} \label{sec3.1: EDP-n}

For a galaxy fitted with a single S\'{e}rsic profile, the disc-like galaxies have a characteristic of $n \sim$ 1, whereas the S\'{e}rsic indices of bulge-dominated galaxies are considered to be $n >$ 2.5 (\citealt{Shen+03, Cebri+2014}). Figure~\ref{fig02-all} shows the dependence of S\'{e}rsic index $n$ and effective radius r$_e$ of GV galaxies on stellar mass in two extreme environments. Dark green squares represent the corresponding median values for GV galaxies in the highest density quarter, while light green squares represent those in the lowest density quarter. The error bars represent the 2$\sigma$ uncertainties drawn from the bootstrap method with the 1000 times resamples.

In the top panels of Figure\ \ref{fig02-all}, we find that the difference in $n$ between the highest and lowest environments is significant at $0.5< z <1$ with the p-value of 2D-KS test $\sim$ 0.003, which reveals that S\'ersic indices ($n$) of GV galaxies are larger in denser environments. It is unlikely to see that GV galaxies are disc-like in the densest environments at $0.5<z <1$, regardless of stellar mass. There are no clear distinctions between two opposite environments at higher redshift ($z>1$) with the corresponding 2D-KS p-value $\gg$ 0.05.
Our result indicates that galaxy environments have a significant impact on galaxy structure at $0.5< z <1$.
This difference in S\'{e}rsic indices between two opposite environments is also reported by \citet{PA+2019}, which found that there is an environmental dependence of S\'ersic indices ($n$) in different stellar mass bins when considering the entire sample including both SFGs and QGs, with denser environments having galaxies with higher S\'ersic indices ($n$) at $z \sim 0.84$.
In addition, there is a correlation between $n$ and redshift: the overall indices $n$ of high-redshift galaxies are smaller than these of low-redshift galaxies. In our sample, it is easier to observe more disc-like GV galaxies at high-redshift.

The bottom panels of Figure~\ref{fig02-all} show the dependence of galaxy size on the environment density and stellar mass. The galaxy size increases by a small amount with stellar mass in given redshift bins, with the Spearman coefficients $>$ 0.3, regardless of the environments. This is consistent with the expectation about the general mass--size relation: more massive galaxies tend to be characterized by a larger radius than their less massive counterparts on average (\citealt{Shen+03, vdW+14}). Our result provides a specific mass--size relation for massive GV galaxies at $0.5<z<2.5$ in two extreme environments. We find no significant difference in size between galaxies in the highest density and lowest density for three redshift bins with KS p-value $\gg$ 0.05, suggesting that a denser environment is not effective enough for inhibiting or stimulating galaxy size.

\begin{figure*}
\centering
\includegraphics[scale=0.8]{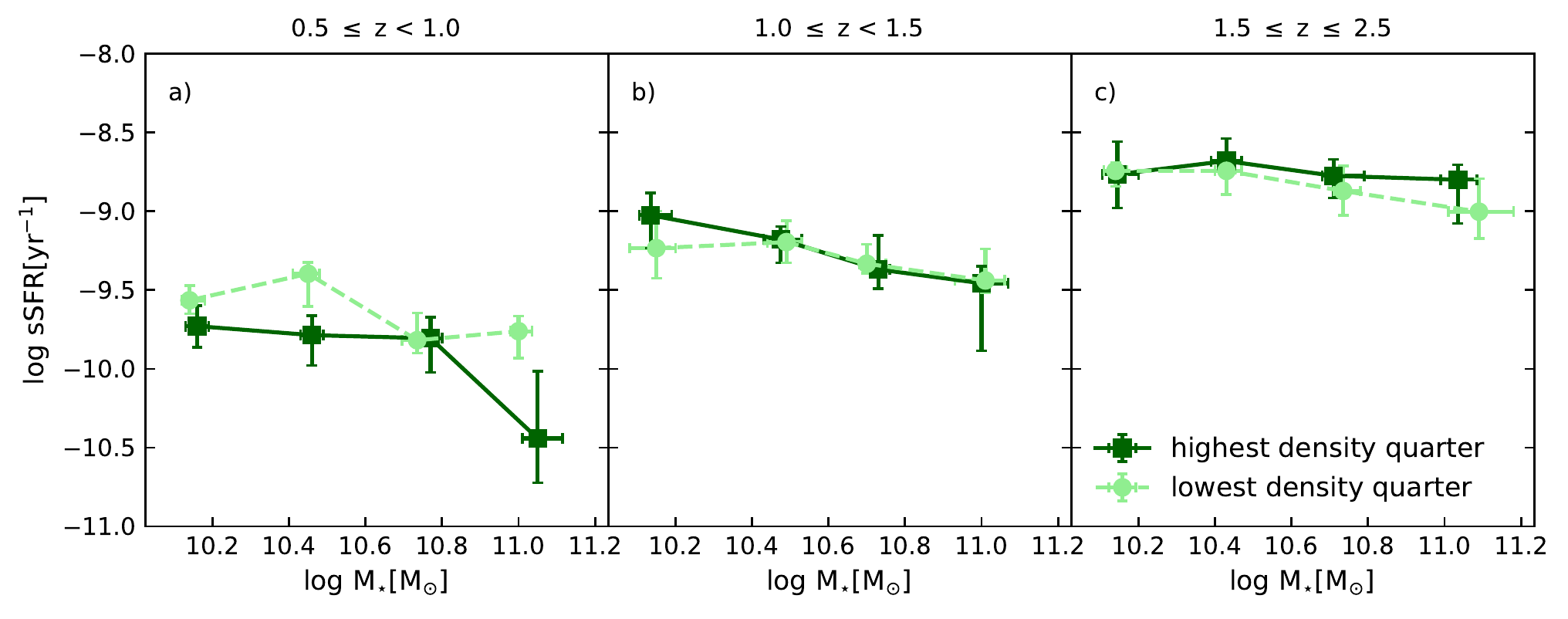}
\caption{Dependence of sSFR for massive GV galaxies on the environment and stellar mass at $0.5 <z <2.5$. The colors and markers are the same as Figure~\ref{fig02-all}.
\label{fig04-sSFR}}
\end{figure*}

\subsection{Non-parametric Structures} \label{sec3.2:NM}

We show the $G$ and $M_{20}$ distributions as functions of stellar mass in different redshift and environment bins in Figure~\ref{fig03-G-M20}. For massive GV galaxies, the average $G$ increases with the cosmic time, while M$_{20}$ generally decreases with decreasing redshifts. A large $G$ value hints that the light distribution of a galaxy tends to be not uniform but concentrated. This could explain why $n$ has a similar redshift evolution as shown in Figure~\ref{fig02-all}. \citet{Peth+16} have approved that the $G$ is sensitive to the $n$. Different from $G$, the decrease of $M_{20}$ is associated with the disappearance of sub-structures in galaxies, such as bars, spiral arms, and bright cores \citep{Lotz+2004}.
In other words, these non-parametric measurements of massive GV galaxies show substantial variations with redshift, i.e, these galaxies at higher redshift tend to have more uniform light distribution globally but more sub-structures locally.
The results of median (or KS) test between the highest and lowest environments show that p-values for $G$ in all redshift bins are larger than 0.05, indicating no environment-related variation. Figure~\ref{fig03-G-M20} also shows the dependence of $M_{20}$ on both stellar mass and at environment overdensity quantiles at $0.5 < z < 1.0$, with 2D-KS p-value $<$ 0.05. The corresponding Spearman coefficients for higher and lower density quarters are about -0.476 and -0.497, respectively, suggesting that $M_{20}$ is negatively dependent on stellar mass. GV galaxies with higher mass tend to have less prominent substrctures.

\begin{figure*}
\centering
\includegraphics[scale=0.8]{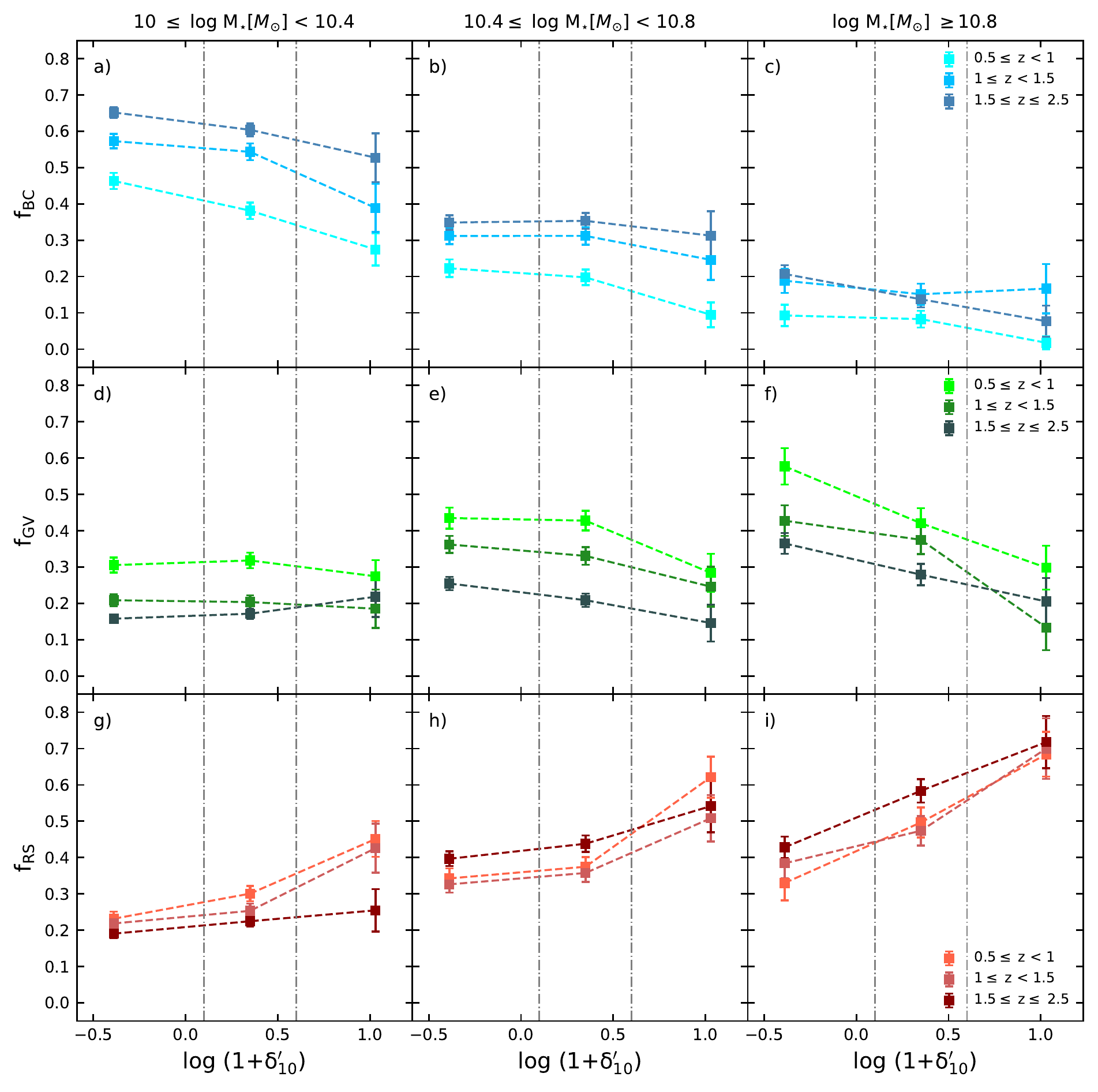}
\caption{The fraction of galaxies in different stellar mass bins as the function of overdensity and redshift for BC (top), GV (middle), and RS (bottom). The dotted line roughly divides the environment into low, medium, and high overdensity. For each panel, colors from light to dark indicate a redshift from low to high for different populations. At a fixed stellar mass and environment, the sum of the fractions of all three populations equals one.
\label{fig_frac_three}}
\end{figure*}

\subsection{Specific Star Formation Rate} \label{sec3.3:SFR}

In Figure~\ref{fig04-sSFR}, we show the sSFR vs. $M_{\star}$ distributions in two extreme environments in three redshift bins. The negative Spearman correlation coefficients ($<$ -0.1) shown in Table\ \ref{tab: corr_KS} reveal that there is a steady drop of sSFR with increasing stellar mass at $0.5 < z < 1$ and $1 < z <1.5$. It is consistent with results of \cite{Bauer+2005}, which reported that sSFR decreases with increasing galaxy stellar mass at $0 < z < 1.5$. It shows the mass dependence for massive GV galaxies, which decreases at high redshift.

In each redshift bin, the dark green squares represent the sSFR--$M_\star$ relation in the highest density quarter, while the light ones are in the lowest density quarter. At $1.0 < z < 2.5$, there is no significant difference of sSFR for GV galaxies in two opposite environments, with the corresponding 2D-KS p-value $>$ 0.05. While at 0.5 $\leq z < 1$, it could be clearly seen the difference of sSFR between the highest and lowest environmental density, where p-value $\sim$ 0.003, especially for the most massive GV galaxies: a sharp drop at the high-mass end for GV galaxies in the highest overdensity.
A possible explanation for this sharp decline is that a denser environment would suppress the star formation activities for more massive GV galaxies at lower redshift. It hints the effect of ``environment quenching''. We would discuss the quiescent fraction in Section~\ref{sec4:FQG} to explain the quenching process more.


\begin{table*}\centering
\ra{1.3}
\tablecaption{Outline of CAE Architecture and Layer Configuration. \label{tab_CAE}}
\begin{tabular}{@{}rrrrcrrrcrrr@{}}\toprule
& \multicolumn{3}{c}{$ 0.5 < z < 1.0$} & \phantom{abc}& \multicolumn{3}{c}{$1.0 < z < 1.5$} &
\phantom{abc} & \multicolumn{3}{c}{$1.5 < z < 2.5$}\\
\cmidrule{2-4} \cmidrule{6-8} \cmidrule{10-12}
& Spearman & Spearman & 2D KS && Spearman & Spearman & 2D KS && Spearman & Spearman & 2D KS\\
& (high) & (low) &  && (high) & (low) &  && (high) & (low) & \\
\midrule

$n$ & 0.292 & 0.244 & 0.003 && 0.056 & 0.084 & 0.567 && -0.046 & -0.031 & 0.528\\

r$_{\rm e}$ & 0.498 & 0.584 & 0.417 && 0.429 & 0.447 & 0.274 && 0.373 & 0.322 & 0.266\\

$G$ & 0.109 & 0.174 & 0.123 && 0.155 & 0.087 & 0.528 && 0.049 & 0.071 & 0.251\\

M$_{\rm 20}$ & -0.476 & -0.497 & 0.015 && -0.214 & -0.155 & 0.407 && -0.137 & -0.081 & 0.755\\

sSFR & -0.258 & -0.275 & 0.003 && -0.327& -0.167 & 0.260 && -0.085 & -0.181 & 0.192\\

\bottomrule
\end{tabular}
\caption{Spearman correlation coefficients and 2-D KS test for physical properties between different environment at $0.5 <z <1.0$, $1.0 < z < 1.5$, and $1.5 <z <2.5$.
\label{tab: corr_KS}}
\end{table*}

\section{Discussion} \label{sec4:FQG}
During the sample selection in Section~\ref{sec2.5: sample select}, we have divided galaxies into three different populations (BC, GV, and RS) according to the corrected rest-frame U-V color. The fraction of each population could give an insight into the quenching timescale, which is defined as the number of each population over the number of the whole sample, i.e., $f_{\rm i} = N_{\rm i}/(N_{\rm BC}+N_{\rm GV}+N_{\rm RS})$, i = BC, GV, and RS. If the quenching timescale is large, BC galaxies will pass through the GV phase slowly and then become quiescent, resulting in a large $f_{\rm GV}$. Otherwise, BC galaxies go through the transition and become quiescent quickly, without a significant rise of $f_{\rm GV}$ (\citealt{Jian+2020}). Therefore, comparing the distributions of $f_{\rm GV}$ as a function of environments with those of $f_{\rm BC}$ and $f_{\rm RS}$ might help us understand the quenching process.

\begin{figure*}
\centering
\includegraphics[scale=0.8]{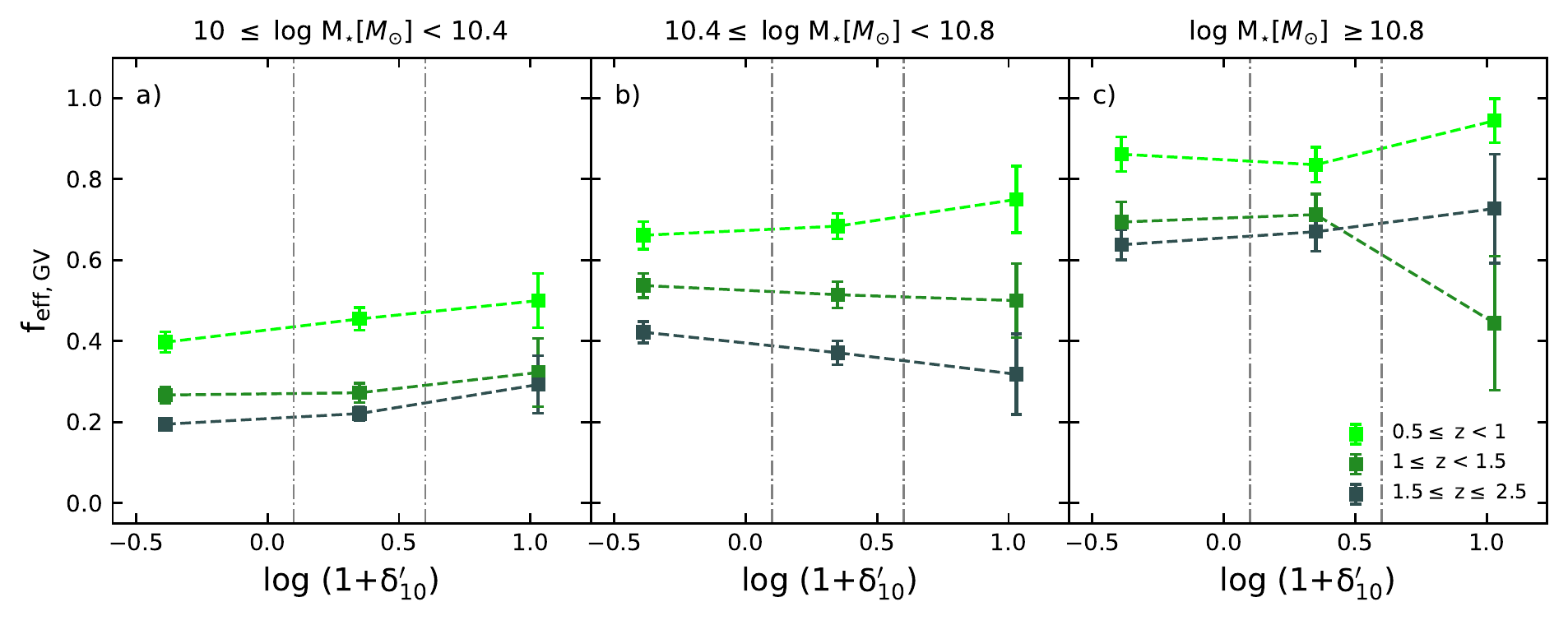}
\caption{The effective fraction of GV galaxies in different stellar mass bins as the function of overdensity and redshift. We define the new effective fraction by normalizing the number of GV galaxies with the number of non-quiescent galaxies.}
\label{fig_f_eff_gv}
\end{figure*}

\subsection{Fractions of Three Populations} \label{sec4.1: frac_three}

Figure~\ref{fig_frac_three} shows the fractions of galaxies belonging to BC (top), GV (middle), and RS (bottom) populations as functions of overdensity in different stellar mass and redshift bins. The vertical dot-dashed lines separate the environment into low ($\log(1+\delta^{'}_{10})< 0.1$), medium ($0.1<\log(1+\delta^{'}_{10})< 0.6$), and high overdensity ($\log(1+\delta^{'}_{10})>0.6$) from the left to right. Regardless of redshifts, $f_{\rm BC}$ keeps a steady decline with the rise of local density, while it is the opposite for the RS fraction, which decreases gradually.

At the low-mass end ($10.0  < \log M_{\star}/M_{\odot} < 10.4$), $f_{\rm BC}$ and $f_{\rm RS}$ are much more dependent on the overdensity compared to $f_{\rm GV}$. While at the high-mass end ($\log M_{\star}/M_{\odot} >$ 10.8), it becomes that $f_{\rm GV}$ and $f_{\rm RS}$ are both sensitive to the overdensity, while $f_{\rm RS}$ is relatively stable. The comparison of $f_{\rm GV}$ in the high- and low-mass bins implies that the quenching timescale of the massive GV galaxies is larger than that of the less massive GV galaxies, because more massive GV galaxies are undergoing the quenching processes, especially in the field at the high-mass end.
So it can reveal the real fact of the influence of stellar mass on the galaxy quenching, called the ``mass quenching''.  

In addition, when we consider the effects of both the environment and stellar mass, it is expected that more massive galaxies in the high overdensity are easier to be quenched into the RS population, resulting in a higher $f_{\rm RS}$ in the denser environment at the high-mass end as shown in the panels f) and i) of Figure~\ref{fig_frac_three}. What is interesting in Figure~\ref{fig_frac_three} is the general pattern of decline in the GV fraction with the increasing overdensity at $0.5<z<1$, even for the less massive galaxies. Meanwhile, galaxies in denser environments have a higher $f_{\rm RS}$, regardless of $M_\star$ and redshifts, which may imply that denser environments have additional effects on the acceleration of quenching. In summary, we find that GV and BC galaxies would be more likely to transform into RS galaxies in the denser environments for all the massive galaxies. Our results also display the effect of stellar mass on the quenching process as previous researches (\citealt{Iovino+2010, PengYJ+10}), thus suggesting that the joint role of stellar mass and environment promotes the quenching process, as evidence of ``mass quenching'' and ``environment quenching'' (\citealt{Vulcani+2012, Darvish+2017}).

\subsection{Effective Fraction of GV Galaxies} \label{sec4.2:Feff}
In general, BC galaxies could be considered as the progenitors of GV galaxies. With the cosmic time, the abundance of RS galaxies and the shortage of BC galaxies naturally leads to the drop-down of GV population in Figure~\ref{fig_frac_three}, especially for galaxies at the high-mass end. Recently, \citet{Jian+2020} have pointed out that the contamination of RS galaxies might bias the relative fraction of GV galaxies. Therefore, the effective fraction, defined as the relative number of GV galaxies over the number of non-quiescent galaxies (i.e., BC and GV galaxies), $f_{\rm eff, GV} = N_{\rm GV}/(N_{\rm BC}+N_{\rm GV})$, can be a better indicator to reflect the galaxy transitional phase more realistically, after removing the effect due to the dominance of the quiescent galaxies.

Based on the new normalized fraction, Figure~\ref{fig_f_eff_gv} shows $f_{\rm eff, GV}$ as a function of overdensity in bins of redshift and stellar mass. After eliminating the RS galaxies, $f_{\rm eff, GV}$ at $0.5< z <1$ in the low overdensity environment is lower than those in the environments with higher overdensity , and the relations between GV fraction and environment reverse compared to those shown in Figure~\ref{fig_frac_three}. This result suggests that the observed negative correlations between $f_{\rm GV}$ and overdensity are mostly due to the reduction of the non-quiescent galaxies with increasing environment densities. The steady increasing trend of $f_{\rm eff, GV}$ with $\log(1+\delta'_{10})$ in the lower redshift bin might support a scenario that environment indeed boosts the beginning of quenching process for SFGs in intermediate-redshift Universe. At $1< z <2.5$, there is no slow upward trend of $f_{\rm eff, GV}$ with the increase of local overdensity, implying that environment has no impact on the beginning of the quenching, although it plays an important role in the complete cessation of the star formation activities as seen in the bottom panels of Figure~\ref{fig_frac_three}.

On the other hand, the overall of $f_{\rm eff, GV}$ at fixed redshift and $\log(1+\delta'_{10})$ significantly increases with $M_\star$, which reveals a strong mass dependence of $f_{\rm eff, GV}$. For more massive GV galaxies at $1<z <1.5$, a sharp drop of the effective fraction of GV from middle to high overdensity is observed, which might be the consequence of the combination of effects from both stellar mass and environment, which is the same as the explains in Section~\ref{sec4.1: frac_three}.

\section{Summary} \label{sec5:Sum}

By utilizing the multi-wavelength data from five 3D-{\it HST}/CANDELS fields, we construct a sample of 7850 massive ($M_{\star} > 10^{10} M_\sun$) galaxies at $0.5< z < 2.5$. Based on the extinction-corrected rest-frame U-V color, we separate the parent sample into BC, GV, and RS galaxies (\citealt{Gu+18, Gu+2019}), resulting in a total number of 2126 GV galaxies. We provide empirical relations for GV galaxies between environment and different physical properties, including $n$, $r_{\rm e}$, $G$, $M_{20}$, and sSFR. We also analyze the fractions of three populations as functions of environments and redshift. Our conclusions are summarized as follows:

1. At $0.5 < z < 1$, GV galaxies have larger S\'{e}rsic indices ($n$) in the denser environment.
The environment seems to have a significant impact on $n$ at $0.5 < z < 1$. It is unlikely to observe disc-like GV galaxies in the densest environment, which suggests that a denser environment would promote the growth of bulge of low-redshift GV galaxies.
We find no significant difference in galaxy size (r$_{\rm e}$) between galaxies in environments with the highest and the lowest overdensity at fixed redshift, indicating that a denser environment is not effective enough to influence galaxy size.

2. Non-parametric measurements of massive GV galaxies show substantial variations with redshift that GV galaxies are more disc-like at higher redshift. Neither stellar mass nor environments have significant impact on $G$ in different redshift bins, while there is a dependence of $M_{20}$ on both stellar mass and environment at $0.5 < z < 1$.

3. At $0.5<z <1.0$, there is a decrease of sSFR from the lowest to the highest environmental density, especially at the high-mass end. A plausible explanation is that a denser environment would suppress the star formation activity of GV galaxies, especially for massive GV galaxies at low redshift.

4. We also discuss the effect of the environment on the fraction for three different populations (BC, GV, RS). At the low-mass end (10.0$<\log M_{\star}/M_{\odot} <$10.4), the BC and RS fractions are much more dependent on the overdensity compared to the GV fraction. While at the high-mass end ($\log M_{\star}/M_{\odot} >$ 10.8), it becomes that the GV and RS fractions are sensitive to the overdensity, and the BC fraction is relative stable. It implies that the quenching timescale of the massive GV galaxies is longer than that of the less massive GV galaxies, which reveals the influence of ``mass quenching''. There is a general pattern of decline in the GV fraction and increase in the RS fraction with the increase of environment density at $0.5< z< 1$. It suggests that denser environments have effects on the quenching process. Considering the effect of mass on three fractions, it could be explained that both denser environments and mass aggregation might promote the galaxy quenching process.

5. The $f_{\rm eff, GV}$ rises gradually with the increase of environmental density, suggesting that environment boosts the beginning of the quenching process. The overall $f_{\rm eff, GV}$ is found to have a positive correlation with $M_\star$, regardless of redshift and environments.
Both original fraction and new effective fraction suggest that stellar mass and environments jointly promote the quenching process.

\acknowledgments
This work is based on observations taken by the 3D-HST Treasury Program (GO 12177 and 12328) with the NASA/ESA HST, which is operated by the Association of Universities for Research in Astronomy, Inc., under NASA contract NAS5-26555.
This work is supported by the Strategic Priority Research Program of Chinese Academy of Sciences (No. XDB 41000000), the National Key R\&D Program of China (2017YFA0402600), the NSFC grant (No. 11973038), and the China Manned Space Project with No. CMS-CSST-2021-A07.
Z.S.L acknowledges the support from China Postdoctoral Science Foundation (2021M700137). Y.Z.G acknowledges support from China Postdoctoral Science Foundation funded project (2020M681281). G.W.F. acknowledges support from Yunnan Applied Basic Research Projects (2019FB007).

\bibliography{reference}{}
\bibliographystyle{aasjournal}

\end{document}